\documentclass[11pt,twoside]{article}

\usepackage{asp2006}
\usepackage{epsf}
\usepackage{lscape}
\usepackage{graphicx}

\begin{document}
\title{Near-IR/optical monitoring programme of the `Arecibo sample of OH/IR stars'}   %%% Fill in title
\author{Francisco Jim\'enez-Esteban,\altaffilmark{1,2} Dieter Engels\altaffilmark{3} and Pedro Garc\'{\i}a-Lario\altaffilmark{4}}   %%% Fill in author names
\altaffiltext{1}{Observatorio Astron\'omico Nacional, Apartado 112, E-28803 Alcal\'a de Henares, Spain}
\altaffiltext{2}{FRACTAL SLNE, Avda San Sebasti\'an 5 3-D, E-38003 Santa Cruz de Tenerife, Spain}
\altaffiltext{3}{Hamburger Sternwarte, Gojenbergsweg 112, D-21029 Hamburg, Germany}
\altaffiltext{4}{Herschel Science Centre. European Space Astronomy Centre, Research \& Scientific Support Deparment of ESA, Madrid, Spain}

\begin{abstract} %%% Abstract to run on from here.

We performed a near-IR/optical monitoring programme from 1999 to 2005
in order to study the variability properties of the `Arecibo sample of
OH/IR stars' (periods, amplitudes, and colour variations). Here we
describe this multi-wavelength long-term monitoring programme. Data
analysis is still in process. Our ultimate goal is to study in
particular the oxygen-rich AGB stars with M\,$>$\,2\,M$_{\odot}$,
which are probably rare among AGB stars discovered optically and/or in
the near-IR, but are common in samples discovered in the mid- to
far-IR.

\end{abstract}

%%% MAIN BODY OF TEXT GOES HERE. CONSULT "INSTRUCTIONS FOR AUTHORS USING
%%% LATEX2E MARKUP", SECTIONS 2.3-2.6 FOR HELP WITH EQUATIONS, FIGURES,
%%% AND TABLES.

\section{The Monitoring Programme}   %%% Top level section head (remove "%" symbol)

Variability properties of AGB stars are still not well known,
specially those of heavily obscured OH/IR stars. A recent study has
shown a diversity of variability properties in OH/IR stars
\citep{Jimenez-Esteban06b}. The number of OH/IR stars studied for
variability is still very low compared to the number of Mira variables
monitored, and neither the upper limit of the period distribution on
the AGB nor the relation between period and other parameters
(amplitude, colour, luminosity) are well defined yet. In addition, the
sequence of colours predicted for oxygen-rich AGB stars with
increasing mass-loss \citep{Bedijn87} is a combination of both
evolution and initial mass effects \citep{Jimenez-Esteban05b}.
Variability properties may be the clue for distinguishing between both
effects.

The so-called `Arecibo sample of OH/IR stars' is the best sample
available to perform this analysis (See \citet{Jimenez-Esteban05a} for
details about this sample).

We performed a 6 years near-IR (JHK) monitoring programme. In total,
we used 156 observing nights distributed in 16 runs at the Calar Alto
1.2\,m telescope (MAGIC) in 1999\,--\,2002, Calar Alto 2.2\,m
telescope (MAGIC) and Tenerife Carlos S\'anchez telescope (CAIN-II) in
2003\,--\,2004, and Calar Alto Spanish 1.5\,m telescope (MAGIC) in
2003\,--\,2005. A subsample of 51 optically bright sources was also
monitored at the optical wavelength (Johnson R) at Hamburg
Oscar-L\"{u}hning-Telescope (CCD) in 2001\,--\,2004.

\section{Preliminary Results} 

We collected in total $\sim$100,000 images which made a high level of
automation desirable. Thus, semi-automated procedures were developed
combining self-written IDL routines with pre-existing routines within
other software packages (IRAF, SEXtractor) to perform the data
reduction and generate one catalogue per individual observation
including the accurate position and aperture photometry of all
point-like sources detected ($>$\,3\,$\sigma$) in the
field. Afterwards, absolute and relative flux calibration was
performed on every catalogue generated. Finally, the photometric data
corresponding to the Arecibo sources were compiled and different epoch
measurements were put together to build up the corresponding light
curves from which we derive the variability properties of each source
in the sample. The results of all these semi-automated procedures are
currently being controlled manually to ensure high reliability on the
final products.

11\,--\,15 photometry points were obtained for each source in the
near-IR, and $\sim$\,20 for the optical monitoring. Typical
photometric accuracy was $<$\,0.2$^{m}$. We have automatically fitted
sinusoidal light curves to the preliminary data and obtained
variability information for 375 OH/IR stars (333 for the first
time). Light curve fits have failed for 57 sources, and for the rest
(276) new periods and amplitudes (255 near-IR, 51 optical) have been
determined. A typical period of 1.3\,$^{yr}$ was found, with a long
tail up to $\sim$\,6\,$^{yr}$ (Fig.1). Less than 5\% of the sample are
non-variable or semiregulars.

%##### IRAS c-c Diagram  #################################
\begin{figure}[!ht]
  \begin{center}
    \rotatebox{-90}{\scalebox{0.25}{\includegraphics{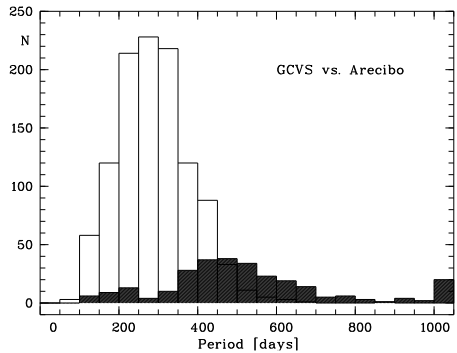}}}
    \caption{Preliminary period distribution of the Arecibo sources
    (shaded) compared to the distribution of the optical Mira variable
    sample included in the GCVS catalogue \citep{Kharchenko02}.}
  \end{center}
\end{figure}
%###################################################

%\subsection{}   %%% Second level section head (remove "%" symbol)
%\subsubsection{}   %%% Lowest level section head (remove "%" symbol)
%\section*{}    %%% Unnumbered top level section head (remove "%" symbol)
%\subsection*{}   %%% Unnumbered second level section head (remove "%" symbol)

%\acknowledgements %%% Text of acknowledgements runs on after this command.

%%% THE BIBLIOGRAPHY
%%%
%%% CONSULT SECTION 3 OF "INSTRUCTIONS FOR AUTHORS" FOR HOW TO USE NATBIB.
%%% AUTHORS ARE ENCOURAGED TO USE EITHER THE "THEBIBLIOGRAPY" ENVIRONMENT
%%% BY UNCOMMENTING (DELETING THE "%" SYMBOL) THE COMMANDS BELOW, OR BY
%%% USING THE BIBTEX ENVIRONMENT. TO FIND OUT WHICH IS APPLICABLE TO YOUR
%%% CONTRIBUTION, CONSULT THE VOLUME EDITORS FOR YOUR PROCEEDINGS.
%%%


\begin{thebibliography}{}

\bibitem[Bedijn (1987)]{Bedijn87}
Bedijn, P.~J.\ 1987, A\&A, 186, 136 

\bibitem[Jim{\'e}nez-Esteban et al. (2006)]{Jimenez-Esteban06b} 
Jim{\'e}nez-Esteban, F.~M., Garc{\'{\i}}a-Lario, P., Engels, D., \& Manchado, A.\ 2006, \aap, 458, 533 

\bibitem[Jim{\'e}nez-Esteban et al. (2005b)]{Jimenez-Esteban05b} 
Jim{\'e}nez-Esteban, F.~M., Garc{\'{\i}}a-Lario, P., \& Engels, D.\ 2005b, AIP Conf.~Proc.~804: Planetary Nebulae as Astronomical Tools, 804, 141

\bibitem[Jim\' enez-Esteban et al.(2005a)]{Jimenez-Esteban05a}
Jim\'enez-Esteban, F.~M., Agudo-M\'erida, L., Engels, D., \& Garc\'{\i}a-Lario, P.\ 2005a, \aap, 431, 779 

\bibitem[Kharchenko et al.(2002)]{Kharchenko02}
Kharchenko, N., Kilpio, E., Malkov, O., \& Schilbach, E.\ 2002, \aap, 384, 925 

\end{thebibliography}
\end{document}